\documentclass{aa} 

\usepackage{graphicx}
\usepackage{txfonts}
\usepackage[english]{babel}
\usepackage{mathtools} 
\DeclareMathOperator*{\argmin}{argmin}

\usepackage{gensymb}
\usepackage{siunitx}

\begin{document}

\title{Supervised detection of exoplanets in high-contrast imaging sequences}

\author{C. A. Gomez Gonzalez
	\inst{1,2}
	\and{O. Absil}
	\inst{1}\fnmsep\thanks{F.R.S.-FNRS Research Associate}
	\and{M. Van Droogenbroeck}
	\inst{3}
}

\institute{STAR Institute, Université de Liège, Allée du Six Août 19c, B-4000 Liège, Belgium
\and{Université Grenoble Alpes, IPAG, F-38000 Grenoble, France}
\and{Montefiore Institute, Université de Liège, B-4000 Liège, Belgium}
}

\date{Received ; accepted }


\abstract  
{Post-processing algorithms play a key role in pushing the detection limits of high-contrast imaging (HCI) instruments. State-of-the-art image processing approaches for HCI enable the production of science-ready images relying on unsupervised learning techniques, such as low-rank approximations, for generating a model PSF and subtracting the residual starlight and speckle noise.} 
{In order to maximize the detection rate of HCI instruments and survey campaigns, advanced algorithms with higher sensitivities to faint companions are needed, especially for the speckle-dominated innermost region of the images. }
{We propose a reformulation of the exoplanet detection task (for ADI sequences) that builds on well-established machine learning techniques to take HCI post-processing from an unsupervised to a supervised learning context. 
In this new framework, we present algorithmic solutions using two different discriminative models: SODIRF (random forests) and SODINN (neural networks). We test these algorithms on real ADI datasets from VLT/NACO and VLT/SPHERE HCI instruments. We then assess their performances by injecting fake companions and using receiver operating characteristic analysis. This is done in comparison with state-of-the-art ADI algorithms, such as ADI principal component analysis (ADI-PCA).} 
{This study shows the improved sensitivity vs specificity trade-off of the proposed supervised detection approach. At the diffraction limit, SODINN improves the true positive rate by a factor ranging from $\sim$2 to $\sim$10 (depending on the dataset and angular separation) with respect to ADI-PCA when working at the same false positive level.} 
{The proposed supervised detection framework outperforms state-of-the-art techniques in the task of discriminating planet signal from speckles. In addition, it offers the possibility of re-processing existing HCI databases to maximize their scientific return and potentially improve the demographics of directly imaged exoplanets.}  



\maketitle

\section{Introduction}

In the last decade, direct imaging of exoplanets has become a reality thanks to advances in optimized wavefront control \citep[for a review see][]{milli16}, specialized coronagraphs \citep{rouan00, spergel01, soummer05, mawet05, kenworthy07}, innovative observing techniques \citep{sparks02, marois06adi} and dedicated post-processing algorithms \citep{lafren07, mugnier09, amara12, soummer12, gomez16}. 
Direct observations of exoplanets provide a powerful complement to indirect detection techniques. They enable the exploration (thanks to their high sensitivity to wide orbits) of different regions of the parameter space, the study of planetary system dynamics, and photometric and spectroscopic characterization of companions. The consensus, after more than ten years of high-contrast imaging, is that massive planets, such as those of HR8799 \citep{marois08hr8799, marois10hr8799}, are rare at wide separations. A meta-analysis of 384 stars conducted by \citet{bowler16} concluded that about $1\%$ of them\footnote{$0.8_{-0.6}^{+1.0}\%$ occurrence rate.} has giant planets at separations between 10 and 1000 AU. On the other hand, from indirect methods, we know that super-Earths and rocky planets are much more common than giant planets. For this reason, the development of new image processing techniques is of key importance for maximizing the scientific return of existing first and second generation high-contrast imaging (HCI) instruments, especially at small separations from the host star. Indeed, the amount of available archival HCI data has increased rapidly with the advent of second generation instruments, such as the Spectro-Polarimetric High-contrast Exoplanet REsearch \citep[VLT/SPHERE,][]{beuzit08} and Gemini Planet Imager \citep[GPI,][]{graham07}. However, the adoption of the latest developments in data management and machine learning in the HCI community has been slow, compared to fields such as computer vision, biology, and medical sciences.

The computational power and data storage increase in the last decade has enabled the emergence of data-driven discovery methods in sciences \citep{ball10}, in parallel to the popularization of machine learning and data science fields of study. Data-driven models are especially important in HCI, if we consider the sheer amount of data that modern high-contrast imaging instruments are producing. Machine learning techniques have proven to be useful in a variety of astronomical applications over the last decade. 
Artificial neural networks are an algorithmic approach proposed a few decades ago in the machine learning community, which is inspired by our understanding of the biology and structure of the brain. Only recently, with graphics processing unit (GPU) computing going mainstream, larger amounts of data, and the use of deep architectures (with increased number of layers and neurons), deep learning has led to breakthroughs in the most challenging areas of machine learning \citep{goodfellow16}. In particular, it has produced impressive results in fields dealing with perceptual data, such as computer vision and language understanding, removing the necessity of hand-crafted features \citep{xie17}. Although neural networks have been used in astronomy since the early nineties \citep{odewahn92, bertin96, tagliaferri03}, the use of deep learning has started to spread only in the last couple of years. Convolutional neural networks \citep[CNN, ][]{lecun89, krizhevsky12} are becoming more and more common for image-related tasks, such as galaxy morphology prediction \citep{dieleman15}, astronomical image reconstruction \citep{flamary16}, photometric redshift prediction \citep{hoyle16}, and star-galaxy classification \citep{kim17}. Other deep neural network architectures, such as autoencoders and generative adversarial networks, have been used for feature-learning in spectral energy distributions of galaxies \citep{frontera17} and for image reconstruction as an alternative to conventional deconvolution techniques \citep{schawinski17}.

\subsection{State-of-the-art image processing techniques for HCI}
A typical HCI planet hunter pipeline includes the production of a science-ready final image, where potential exoplanets are flagged by visual inspection aided by the computation of a signal-to-noise (S/N) metric. In this study, we adopt the S/N definition of \citet{mawet14ss} which addresses the small sample statistics effect at small separations. In the case of angular differential imaging \citep[ADI,][]{marois06adi} data, the generation of a final image usually relies on differential imaging post-processing techniques. The purpose of these techniques is to reduce the image dynamic range, by modeling and subtracting the contribution from the high-flux pixels belonging to the residual starlight and from the quasi-static speckle noise. This procedure, also called model PSF subtraction\footnote{Here we define the model PSF as the algorithmically built image that we use with differential imaging techniques for subtracting the scattered starlight and speckle noise pattern in order to enhance the signal of disks and exoplanets.}, produces residual final images where, unfortunately, part of the companion signal is lost due to it being fitted in the model PSF (companion self-subtraction). Among the model PSF subtraction techniques, we count LOCI \citep{lafren07}, principal component analysis (PCA) based algorithms \citep{soummer12,amara12}, and LLSG \citep{gomez16}. All these approaches use different types of low-rank approximation to generate a model PSF. A different approach is taken by ANDROMEDA \citep{mugnier09,cantalloube15}, which employs maximum likelihood estimation on residual images obtained by pairwise subtraction within the ADI sequence.

The exoplanet detection problem is critical as it triggers all subsequent steps, such as the determination of position, flux and other astrophysical parameters (characterization) of potential companions. The task of detecting potential companions with model PSF subtraction techniques lacks automation. It boils down to the visual identification of patches of pixels sharing the same properties, such as bright regions on the images, and resembling the instrumental PSF. Therefore, the detectability of significant blobs by visual inspection is limited by human perception biases. This process is aided by the computation of the S/N metric, but computing S/N maps is ultimately upper bounded by the performance of the chosen model PSF subtraction technique. Moreover, the S/N metric does not deal with the truthfulness of potential companions. Other approaches to detecting blobs such as the Laplacian of Gaussian and the matched filtering \citep{ruffio17} suffer from the same problem. For a review of general purpose source detection techniques on astronomical images, see \citet{masias12}. 

Advanced approaches with higher sensitivities to dim companions are needed, especially for the speckle-dominated innermost region of the images. Such approaches must address the issues of the visual vetting and S/N map computations by producing per-pixel likelihoods or probabilities of companion presence for a given ADI sequence. The maximum likelihood approach of ANDROMEDA, while a step in this direction, has not been thoroughly benchmarked against state-of-the-art approaches. Comparative contrast curves show its performance to be at the same level as full-frame ADI-PCA \citep{cantalloube15}. 

A different approach to detecting exoplanets through HCI is the use of discriminative models, as it has been proposed by \citet{fergus14} for the case of multiple-channel SDI data. The DS4 algorithm, an extension of the S4 algorithm, adopts a discriminative approach based on support vector machines trained on a labeled dataset. This dataset is composed of negative samples taken directly from the input data and positive samples generated by injecting synthetic companions. Unfortunately, there is no publication describing the details of this algorithm or robustly assessing its performance (see the discussion section of \citet{fergus14}). 

\begin{figure*}
	\centering
	\includegraphics[clip,trim=0 140 250 0, width=17cm]{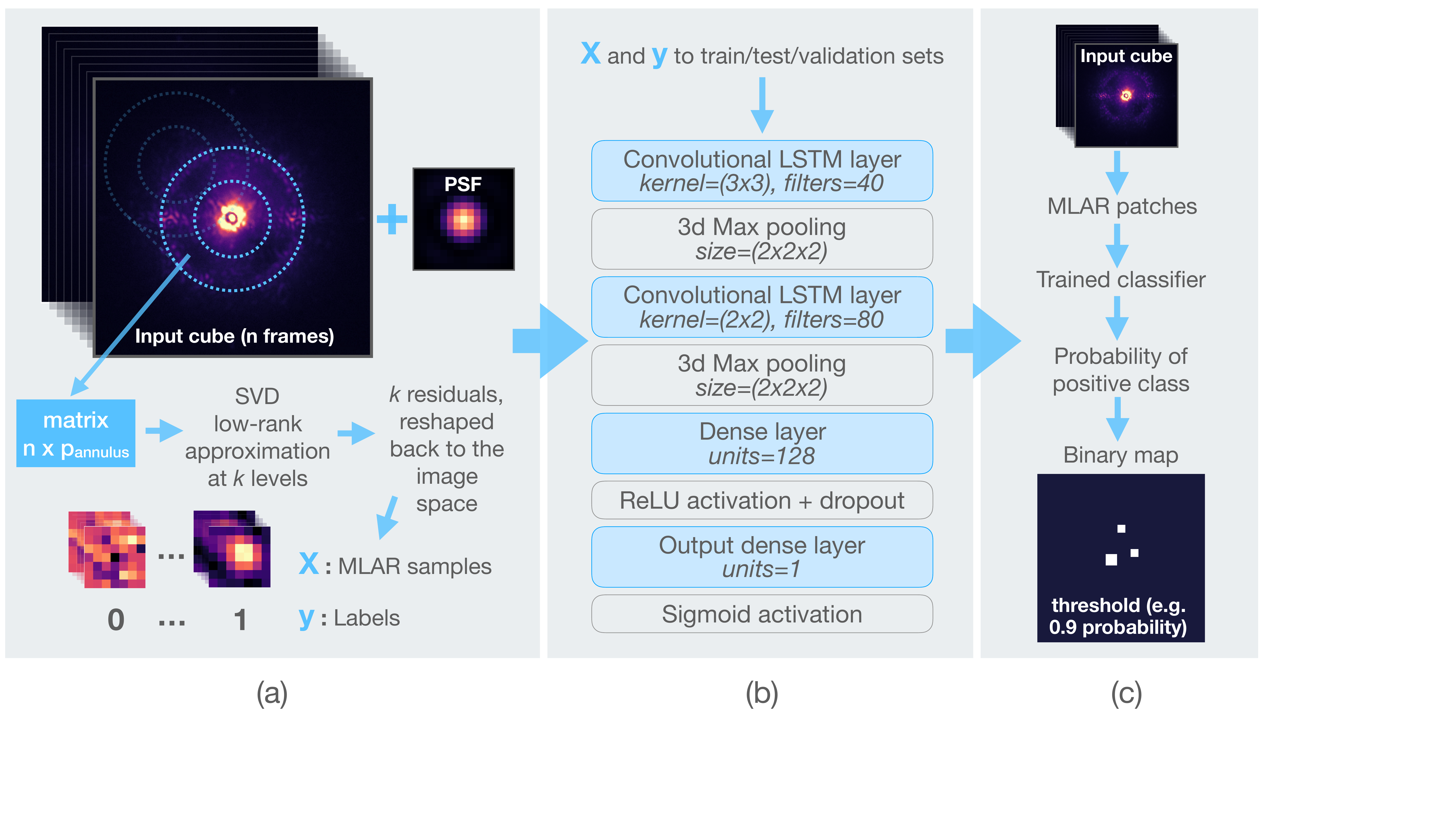}
	\caption{The three stages of our supervised detection framework. Panel (a) illustrates the labeled data generation step. The ADI sequence and off-axis PSF template are examples of VLT/SPHERE data. Panel (b) illustrates the model training step for the case of SODINN. SODIRF uses a random forest classifier instead of a deep neural network. Panel (c) concerns the evaluation of the trained model on the original cube and shows the schematic representation of the output detection map.}
	\label{fig:odinn_scheme}
\end{figure*}

\subsection{From unsupervised to supervised learning} \label{sec:unsuptosup}
Differential imaging post-processing approaches rely on unsupervised learning techniques, such as low-rank approximations, to enable the production of final residual images. The detection ability of these techniques 
depends on a variety of factors, including the number of frames in the sequence, the total range of field rotation, the distance of a companion to its parent star, the companion flux with respect to the star, and the aggressiveness of the differential imaging subtraction approach. 

Our approach here consists in a reformulation of the exoplanet detection task as a supervised binary classification problem. Supervised learning uses a considerable amount of labeled data (or ground truth) in order to train a discriminative model and produce predictions. Depending on the model used, two algorithms are proposed: SODIRF, which stands for Supervised exOplanet detection via Direct Imaging with Random Forests, and SODINN, which stands for Supervised exOplanet detection via Direct Imaging with deep Neural Networks.

The first stage or our method addresses the challenge of generating a large labeled dataset from a single ADI image sequence. As we show in Sec. \ref{sec:labeldata} this procedure relies on the injection of synthetic companions and a technique called data augmentation, which is widely used in deep learning. Once our model is trained on this labeled dataset, it can be applied to the input ADI sequence for evaluation without risk of overfitting\footnote{Model overfitting occurs when a machine learning algorithm models random noise in the labeled training data, limiting the prediction power on unseen new data (lack of generalization). For high-capacity models, such as deep neural networks, overfitting also occurs when the model memorizes the labeled training data limiting the prediction ability on new data samples.}. Figure~\ref{fig:odinn_scheme} shows a diagram of our novel framework for the case of SODINN.

The fact that SODIRF and SODINN can be trained on a labeled dataset created from a given ADI sequence means that these models are fine-tuned to each ADI sequence \citep{braham16}. We have tested SODIRF and SODINN on coronagraphic ADI sequences from different instruments. To validate our results, we focus on two datasets (one of them with a known companion) that are very different in terms of their characteristics. The first dataset, an $L'$ band VLT/NACO sequence on $\beta$~Pic \citep{oabsil13} and its companion \citep{lagrange10}, consists of 612 frames with 8 sec of effective integration time, and has a total field rotation of 83 degrees. This $\beta$~Pic dataset is described in \citet{oabsil13} along with the pre-processing procedures applied to generate the calibrated cube (or reduced image sequence) that we used here. The second dataset is a VLT/SPHERE sequence on V471 Tau \citep{hardy15} acquired with the Infra-Red Dual-band Imaging and Spectroscopy \citep[IRDIS,][]{Dohlen_2008_IRDIS} subsystem. It consists of two sequences (in the H2 and H3 bands\footnote{The SPHERE/IRDIS instrument provides dual-band imaging thanks to the use of a beam splitter located downstream the coronagraphic mask.}) with 50 frames, each one with 64 sec of integration time, and a total field rotation of 30 degrees. The pre-processing steps applied to the V471 Tau dataset are described in \citet{hardy15}. Throughout this study, and for simplicity, we assume that (1$\times$)${\rm FWHM} = 1\lambda/D =4$\,pxs.

This paper is organized as follows. In Sec. \ref{sec:labeldata}, we describe our labeled data generation strategy for ADI datasets. Section \ref{sec:discrmodel} describes the two proposed classification approaches using random forests and deep neural networks. Section \ref{sec:predstage} explains the prediction stage of our supervised detection approach. Section \ref{sec:perf_eval} presents our performance assessment study using signal detection metrics for comparing SODIRF and SODINN to state-of-the-art HCI algorithms, and Sec. \ref{sec:conclusions} presents the conclusions.

\section{Generation of a labeled dataset} \label{sec:labeldata}
The generation of a labeled dataset requires a transformation of the ADI image sequence that suits better a supervised learning problem and enables us to create examples of two distinguishable classes: one representing the companion signal and the other the speckles and background areas. Therefore, we work on patches, instead of full frames, in order to get a different view of the image sequence. This choice is motivated by the fact that the exoplanet's signal spatial scale is small compared to the frame size, and that it  facilitates the creation of a large labeled dataset even from a single ADI sequence, as explained hereafter. 

Working with 2D patches directly from a pre-processed ADI sequence does not facilitate the generation of two distinguishable classes. This is mainly due to the high dynamic range caused by the presence of residual starlight. Our initial tests using 2D patches were not successful and this motivated the use of a different view of the data.
Our labeled dataset is composed of 3D residual patches, at several Singular Value Decomposition (SVD) approximation levels, hereafter referred to as Multi-level Low-rank Approximation Residual (MLAR) samples. They can be understood as computing annulus-wise PCA residual patches at different numbers of principal components (PC). Working with these MLAR patches, we replace the ADI temporal information with the patch evolution as a function of the approximation level. 

The MLAR samples are built in the following way. Consider a matrix $M \in \mathbb{R}^{n\times{p}}$ whose rows contain the pixels inside a centered annulus of a given width. $n$ is the number of frames in the ADI sequence and $p$ is the number of pixels in the given annulus. Recall that singular value decomposition (SVD) is a matrix factorization such that:
\begin{equation}
M=U\Sigma V^{ T}=\sum _{i=1}^{n}{ \sigma_{i}u_{i}v_{i}^{T}  },  
\end{equation}
where the vectors $u_{i}$ and $v_{i}$ are the left and right singular vectors, and $\sigma_{i}$ the singular values of $M$. SVD is involved in several least-squares problems, such as finding the best low-rank approximation of $M$ in the least-squares sense, i.e.,
\begin{equation}
\argmin_{X}{\left\|M-X\right\|}_{F}^{2}, 
\end{equation}
where $\left\| \cdot \right\|_{F}^{2}$ denotes the Frobenius norm. By keeping $k$ right singular vectors, we form a low-dimensional subspace $B$ capturing most of the variance of $M$. The residuals are obtained by subtracting from $M$ its projection onto $B$:
\begin{equation}
	R = M-MB^TB.
	\label{svd_res}
\end{equation}
This residual matrix is later reshaped to the image space, de-rotated and median combined as the usual ADI workflow dictates. In general, the larger the value of $k$, the better the reconstruction and the smaller residuals (with less energy or standard deviation).

Instead of choosing one single $k$ value for estimating the low-rank approximation of $M$ and obtaining a single residual flux image (which is the goal of PCA-based approaches), we choose multiple $k$ values sampling different levels of reconstruction. The MLAR patches are obtained by cropping square patches, of odd size and about twice the size of the FWHM, from the sequence of final residual frames obtained for different $k$. Defining the values of $k$ relies on the cumulative explained variance ratio (CEVR). Let $\hat{M}$ be the matrix $M$, from which its temporal mean has been subtracted, and $\hat{\sigma_i}$ the singular values of $\hat{M}$. The explained variance ratio for the $k^{\rm th}$ singular vector is defined as:
\begin{equation}
	\frac{(\hat{\sigma_k}^2/n)}{\sum_i{\hat{\sigma_i}^2}},
	\label{eq:expvar}
\end{equation}
where $i$ goes from one to $min(n,p)$. It measures the variance explained by each singular vector and the CEVR measures the cumulative explained variance up to the $k^{\rm th}$ singular vector. Sensible values for $k$ lie within the interval from 0.5 to 0.99 CEVR (for one example, see left panel of Fig.~\ref{fig:cevr_flux}), but depend on each particular dataset. The number of steps in this interval can be tuned, although the general rule is that more steps in the MLAR patches lead to more expressive samples that generally lead to higher classification power and a better discriminative model. In our tests, with 8 to 20 approximation levels, we could train models with outstanding accuracy. 

By using this data transformation, we are able to generate MLAR samples from our two classes, one containing the signature of a companion (\emph{positive} class $c^+$) and the other representing the background and speckle diversity (\emph{negative} class $c^-$). Each sample has an associated label $y \in \left \{c^-, c^+ \right \}$. 

\subsection{Generation of the $C^+$ MLAR samples}
\begin{figure}
	\centering
	\includegraphics[width=9.2cm]{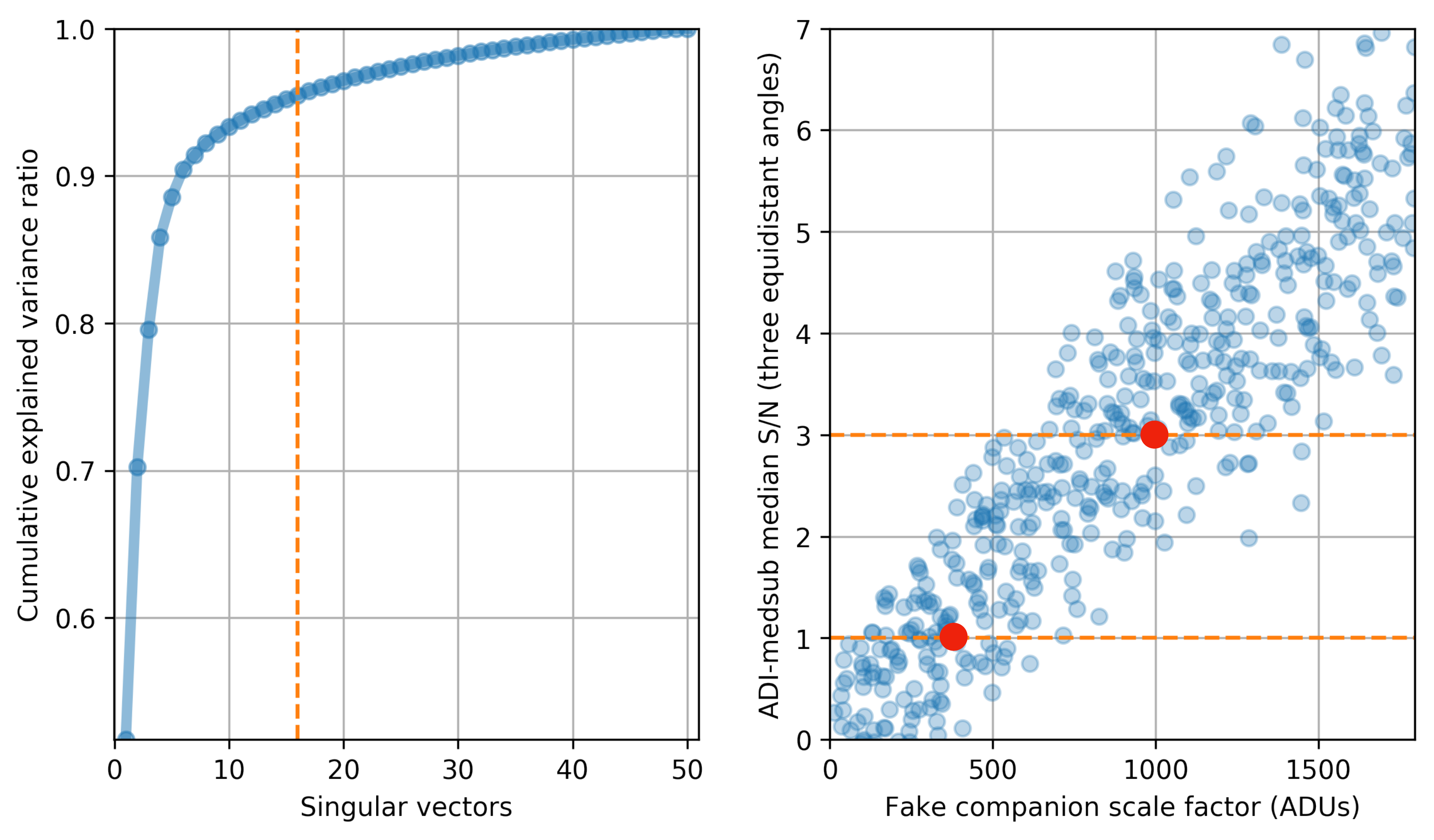}
	\caption{Generation of a labeled dataset. The left panel illustrates the procedure for determining the approximation levels and shows the cumulative explained variance ratio as defined by Eq. \ref{eq:expvar}. The vertical dotted line is located at the maximum number (16) of singular vectors used in this case. The right panel illustrates the determination of flux intervals and shows the median S/N of injected companions, in an ADI-median subtracted residual frame, as a function of the scaling factor. The red dots denote the lower and upper bounds of the companion injections for generating MLAR samples of the \emph{positive} class.} 
	\label{fig:cevr_flux}
\end{figure}

The creation of the \emph{positive} class relies on injecting an off-axis PSF template, a procedure accepted within the HCI community for generating synthetic data and assessing the sensitivity limits of image processing algorithms and instruments. The PSF template is usually obtained from observations of the same star during the same observing run and without a coronagraph. The injection consists in the addition of such PSF template (on each frame of the sequence) at a given location with a random brightness from a predefined interval. This interval must be carefully chosen to avoid class overlap, which occurs when a same MLAR sample (or very similar) appears as a valid example of both classes. This can happen when the lower bound of our brightness interval (or planet to star contrast) is too low, in which case the signature of the companion signal is hardly distinguishable from the one of the background polluted with quasi-static speckles. Sensible lower and upper bounds can be estimated in a data-driven fashion by injecting fake companions and measuring their S/N in residual frames obtained through classical ADI median subtraction. Fluxes leading to S/Ns in the interval [1,3] are usually good for our purpose (see right panel of Fig.~\ref{fig:cevr_flux}). These flux intervals are defined in an annulus-wise fashion and are therefore related to the radial flux profile of the images.

\subsection{Generation of the $C^-$ MLAR samples}
The generation of the samples from the \emph{negative} class, representing everything but the signal of companions (the background and speckles), relies on the exploitation of the rotation associated to an ADI sequence and common machine learning data augmentation techniques\footnote{This refers to the process of creating synthetic data and adding these to the training set in order to make a machine learning model generalize better (see section 7.4 of \citet{goodfellow16}).}. The generation of a large number of \emph{negative} samples faces two main difficulties. First, the fact that with a single ADI image sequence, we obtain a single realization of the residual noise (in a PCA-based differential imaging context). Second, the number of patches we can grab from a given 1$\times$FWHM annulus is orders of magnitude smaller than the number of samples that are needed in the labeled dataset. If we feed these samples to a classifier, it would quickly memorize them, and that would produce strong overfitting (especially in the case of a deep neural network). Our dedicated data augmentation process addresses these issues and can be summarized by the following steps: 
\begin{enumerate}
	\item We randomly grab MLAR patches (as explained at the beginning of Sec. \ref{sec:labeldata}) centered on up to ten percent of the pixels in a given annulus. Optionally, a chosen region (circular aperture) of the ADI frame sequence can be masked to conceal a known, true companion. The corresponding patches are then ignored.
	\item We flip the sign of the parallactic angles when derotating the residual images (after reshaping to image space the residuals obtained in Eq. \ref{svd_res}) to obtain final median combined images that preserve the noise correlation and keep the same statistical properties, while blurring any astrophysical signal. We grab all the available MLAR patches from the given annulus.
	\item We randomly pick groups of three samples from the two previous subsets and average them to produce new samples.
	\item Finally, we perform random rotations and small shifts of the MLAR samples obtained in the previous three steps to	create even more diversity. The same rotation angle and shift is applied to all the slices of a given MLAR sample.
\end{enumerate}

\begin{figure}
	\centering
	\includegraphics[width=9cm]{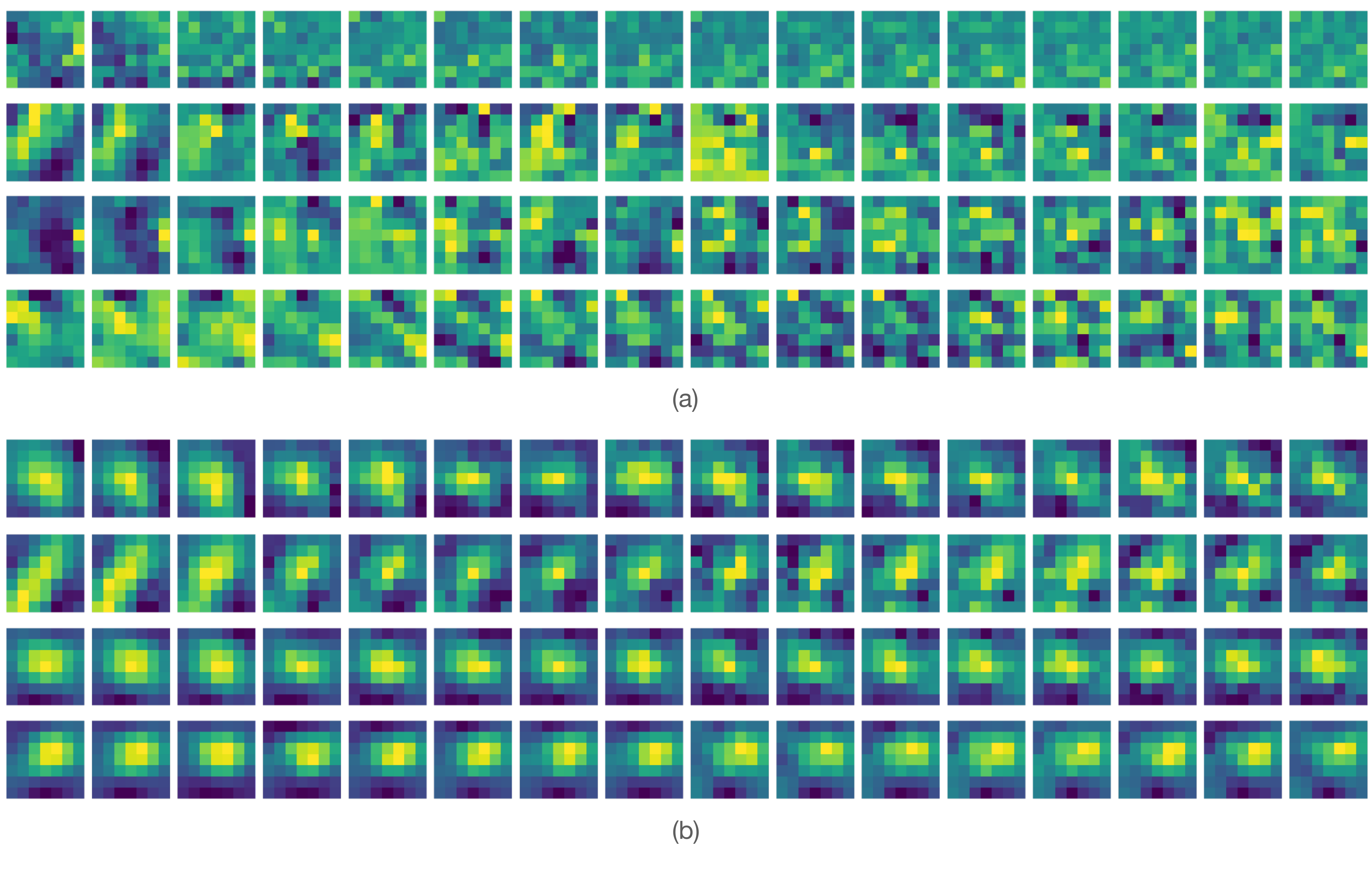}
	\caption{MLAR samples from the \emph{positive} and \emph{negative} classes obtained with up to 16 singular vectors. The CEVR for these MLAR samples are shown in Fig.~\ref{fig:cevr_flux}. (a) Each row corresponds to a random MLAR sample from the \emph{negative} class (background  and speckles). (b) Each row corresponds to a MLAR sample from the \emph{positive} class (exoplanet signal). The \emph{positive} samples are shown, from top to bottom, with increasing flux. Every slice of the MLAR sample is normalized in the interval [0,1]. }
	\label{fig:patches}
\end{figure}

In the end, the $C^+$ MLAR samples contain the signature of the injected companions and the $C^-$ MLAR samples contain augmented samples without companion signal. Thanks to this strategy, we avoid showing the samples from the original ADI sequence to our classifiers, thus reducing model overfitting. Note that the pixel values in each slice of the MLAR sample are normalized in the interval [0,1], bringing all the labeled dataset to the same value range. In panels (a) and (b) of Fig.~\ref{fig:patches}, we show a few examples of the resulting MLAR samples composing our labeled data set. The patch size was set to seven pixels. The MLAR \emph{positive} samples shown in panel (b) clearly illustrate the exoplanet PSF morphological distortion introduced by differential imaging post-processing, as a function of the aggressiveness (analogous to the number of PCs used in a PCA-based post-processing approach). This is related to the well-known problem in HCI of companion self-subtraction. The PSFs of the companions clearly degrade as the CEVR increases (they eventually vanish when $k$ is close to $min(n,p)$), which affects the positions of the PSF centroids.

We use the \texttt{VIP} \texttt{Python} library \citep{gomez17} for low-level image operations and the generation of labeled datasets. The calculations for producing the MLAR samples are done on CPU in a parallelized way and the SVD computations use the randomized SVD algorithm proposed by \citet{halko11} to decrease the computation time. We use the above described procedure to generate a balanced labeled dataset of several hundreds of thousands MLAR samples (with the same amount of $c^-$ and $c^+$ samples). Here again the general rule is that more samples are better for the discriminative power of our models. A thorough analysis of the influence of the labeled dataset size on the performance of our discriminative models has yet to be performed.

\section{Discriminative model} \label{sec:discrmodel}
The fact that the footprint of a companion in the MLAR patches is different from the one of a speckle or a background area enables the formulation of the exoplanet detection as a binary classification task. The role of the discriminative model, in the proposed supervised detection framework, is to disentangle the exoplanet signal signature $c^+$ from the background and speckle pattern $c^-$. The classifier achieves this by learning a mapping from the input MLAR samples to their corresponding labels. Once the model is trained, it is able to make predictions $\hat{y} \in \left \{ c^-, c^+ \right \}$ on new samples. The probabilistic classifiers we discuss in this study assign to each sample a confidence score of class membership, which we call probability hereafter, from which we obtain a class prediction by applying a threshold of 0.5. In the following Sections, we propose two ways of approaching the classification step, one using random forests (SODIRF) and a more sophisticated one using deep neural networks (SODINN). In Sec. \ref{sec:perf_eval} we focus on the confidence scores provided by SODIRF/SODINN and explore different probability thresholds using signal detection theory metrics suited for performance assessment of binary probabilistic classifiers.

\subsection{Random forest based approach}
A random forest \citep{breiman01} is a type of ensemble learning model. Ensemble methods rely on the introduction of random perturbations into the learning procedure (of the mapping function) in order to produce several different models from a single labeled dataset, and the combination of the predictions of those models to form the prediction of the ensemble. In particular, a random forest fits a multitude of decision trees on various bootstrap sub-samples of the labeled dataset, and performs averaging of their probabilistic predictions to improve the predictive accuracy of the model (by reducing the variance of the ensemble if compared to single decision tree).
A detailed description of the random forest algorithm is beyond the scope of this paper. For details we refer the reader to \citet{louppe14}. In the case of SODIRF, we must create a 2D matrix of samples versus features (the pixels of each MLAR sample) suitable for training the random forest classifier. This feature matrix is constructed by vectorizing the MLAR samples and stacking them in a matrix. SODIRF is implemented using the \texttt{scikit-learn} \texttt{Python} machine learning library. This implementation of a random forest combines the decision tree classifiers by averaging their probabilistic prediction. SODIRF uses 100 fully developed trees to form the ensemble model and a simple train-test splitting procedure for the training stage. The random forest model achieves a good test accuracy (over 99.5\%).

Random forests can be efficiently trained on CPUs, in just a few minutes, exploiting modern multi-processor architectures, unlike deep neural networks (such as deep CNNs), which require last generation GPUs and more computing time to be trained. The models differ not only in terms of the computational cost but also in terms of performance, as we show in Sec. \ref{sec:perf_eval}.

\subsection{Deep neural network based approach}
Deep learning is a particular subfield of machine learning that relies on the use of successive layers of representations, enabling the creation of models with high levels of abstraction. Deep neural networks are a particular kind of artificial neural network architecture that learn these layered representations by stacking many layers of neurons one after the other. CNNs are a type of deep learning model for processing data having a grid-like topology (e.g. images), and are almost universally used in computer vision. CNNs are also employed for processing time series and 3D input data. On the other hand, recurrent neural networks \citep[RNN, ][]{rumelhart86} are a powerful type of neural network particularly designed for sequence modeling. Long-short term memory \citep[LSTM, ][]{hochreiter97} networks are a kind of RNN widely used in machine translation, large-vocabulary speech recognition, and text-to-speech synthesis, thanks to its ability of learning long term dependencies.

SODINN makes use of deep neural networks to exploit the 3D structure of the MLAR samples. We have explored two different types of networks suited for learning spatio-temporal (3D) dependencies: 3D convolutional networks \citep{tran15} and convolutional LSTM networks \citep{shi15}. By using these network architectures, we can directly feed the model with the MLAR samples thereby preserving their 3D structure, as opposed to SODIRF. In order to find a model with the best sensitivity vs specificity trade-off, we have performed a manual search to explore combinations of the two architectures and different hyperparameters. We obtain the best results with convolutional LSTM layers, combining convolutional and LSTM architectures, and we choose it for building SODINN's classification model.

As shown in Fig.~\ref{fig:odinn_scheme}, SODINN's classifier architecture consists of two convolutional LSTM layers, the first with 40 filters of size 3$\times$3 and the second with 80 filters of size 2$\times$2. Each convolutional LSTM layer is followed by a max pooling layer \citep{boureau10} which aggregates the activations of neighboring units by computing the maximum of 2$\times$2$\times$2 3D patches. The network follows with a fully connected layer featuring 128 hidden units. A rectified linear unit \citep[ReLU, ][]{nair10} activation (non-linearity) is applied to the output of the dense layer and a  dropout \citep{hinton12, srivastava14} regularization is applied to the resulting activations. Finally, the output layer of the network is a sigmoid unit. The network weights (\num{2.5e5} to \num{1e6} learnable parameters depending on the size of the FWHM) are initialized randomly using a Xavier uniform initializer and are learned by back-propagation with a binary cross-entropy cost function:
\begin{equation}
	\mathcal{L} = -\sum_n (y_n\ln(\hat{y}_n) + (1-y_n)\ln(1-\hat{y}_n)),
\end{equation}
where $y_n$ is the true label of the $n^{th}$ MLAR sample and $\hat{y}_n = p(c^+ \mid \rm MLAR \; sample)$ is the probability that the $n^{th}$ MLAR sample belongs to the \emph{positive} class. The architecture of the neural network is not dataset dependent.

The labeled data is divided into train, test (ten percent of the initial labeled samples), and validation sets. The optimization of deep networks, with a large number of parameters, is accomplished with mini-batch stochastic gradient descent. It works by drawing a random batch from the training set, performing a forward pass (running it through the network) to obtain predictions $\hat{y}$, computing the loss score on this batch, and the gradient of the loss with regard to the parameters of the network (which is called a backward pass). The parameters or weights are then changed in the direction opposite to the gradient \citep{chollet17}. The aim of this process is to lower the loss on the batch by a small step, also called learning rate. The whole process of learning the weights (that minimize the loss) is made possible by the fact that neural networks are chains of differentiable tensor operations. Therefore it is possible to use the \emph{backpropagation} method, by applying the chain rule of derivation to find the gradient function mapping the current parameters and current batch of data to a gradient value.

We adopt the Adam optimization strategy \citep{kingma14}, which extends classical stochastic gradient descent and computes individual adaptive learning rates for different parameters from estimates of first and second moments of the gradients. We use a step size of 0.003 and mini-batches of 64 training samples. We include an early stopping condition monitoring the validation loss. Usually, our model is trained with 15 epochs (passes of the stochastic gradient descent optimizer through the whole train set) reaching 99.9\% validation accuracy. SODINN's neural network classifier is implemented using the highly modular and minimalist \texttt{Keras} library \citep{chollet15keras} using its \texttt{Tensorflow} \citep{tensorflow15} backend. The model is trained on a NVIDIA DGX-1 system using one of its eight P100 cards in about one hour. Training such network is possible on any computer with a dedicated last generation GPU, such as a NVIDIA TitanX. Almost the same runtime is achieved when training the model on a much cheaper GTX 1080 Ti card installed on a conventional server.

\begin{figure}
	\includegraphics[clip,trim=25 10 25 0, width=9cm]{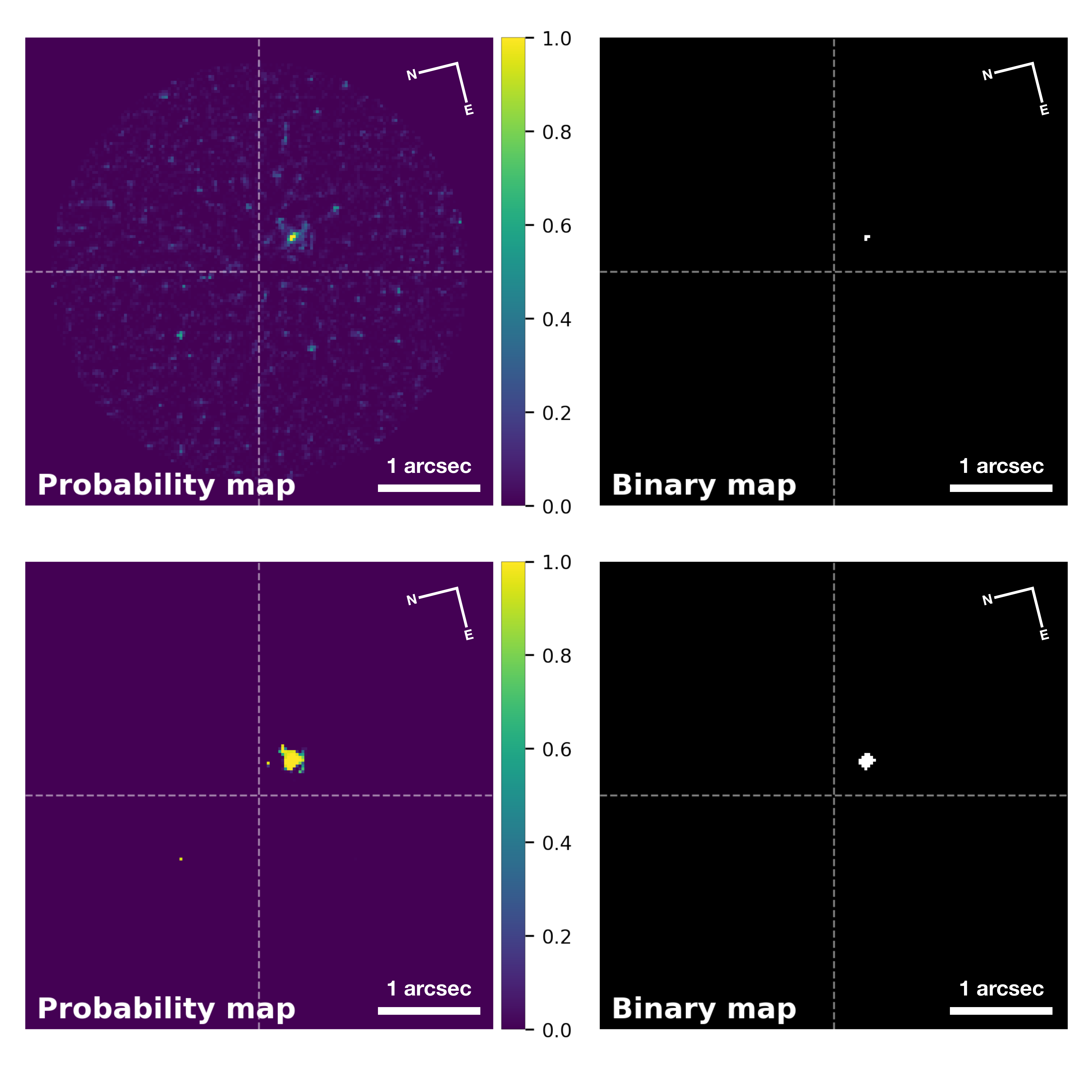}
	\caption{SODIRF and SODINN outputs for the VLT/NACO $\beta$~Pic dataset. Top panels show SODIRF's probability (left) and binary detection maps (right). Bottom panels correspond to SODINN's output. Both binary detection maps are obtained with a 99\% probability threshold.}
	\label{fig:pred_betapic}
\end{figure} 

\section{Prediction stage} \label{sec:predstage}
Once the models are trained, they are applied to the input data cube. First, we perform the same transformations (same CEVR intervals) to the input ADI sequence to obtain MLAR patches centered on each one of the pixels of the frame. The discriminative model then classifies these MLAR patches, assigning a probability of membership to the \emph{positive} class, $p(\hat{y}=c^+ \mid \rm MLAR \; sample)$. For SODINN, the prediction stage is just a forward pass of a given test sample through the trained deep neural network to produce an output probability. In our supervised framework, grabbing MLAR patches for each pixel of the frame, enables the estimation of a class probability in a detection map. This map is then thresholded at a desired level of $c^+$ class probability. The probability and binary maps are the outputs of both SODIRF and SODINN, as exemplified in Fig.~\ref{fig:pred_betapic} for the VLT/NACO dataset. We can see how the binary maps clearly reveal the presence of $\beta$~Pic b, without false positives, for this probability threshold.

For comparison, in a differential imaging PCA-based approach, one would tune the number of PCs that works best for a companion at a given radial distance and obtain a residual flux image. This trial and error process leads to a single realization (using one $k$ value) of the residuals, which is then visually inspected to identify companions or is turned into a S/N map. In the case of our supervised detection method, the predicted probability (or detection criterion) is evaluated independently for each pixel on the frame and does not suffer from the small sample statistics issue or, for that matter, human perception biases. This is a huge improvement compared to differential imaging where the S/N metric requires to take into account the annulus-wise noise at the separation of a given test resolution element. 

In order to test the validity of our training procedure, we injected faint fake companions in the ADI sequence used to generate the labeled dataset, without masking the injected companions, to simulate the situation when we face a new dataset with real unknown exoplanets. Afterwards, we checked whether the trained models were able to detect these pre-existing companions. In this test, the injected companions could be recovered with a high success rate, which demonstrates that our approach prevents overfitting at the labeled dataset generation stage. Therefore, we conclude that our framework can be safely applied to new ADI datasets and the performance assessment shown in Sec. \ref{sec:perf_eval} is fair. We would like to emphasize that having access to multiple datasets taken with the same instrument (survey data), would enable training a more general model and would depend less strongly on the proposed data augmentation procedure. 

\begin{figure*}
	\centering
	\includegraphics[clip,trim=10 60 10 0, width=18cm]{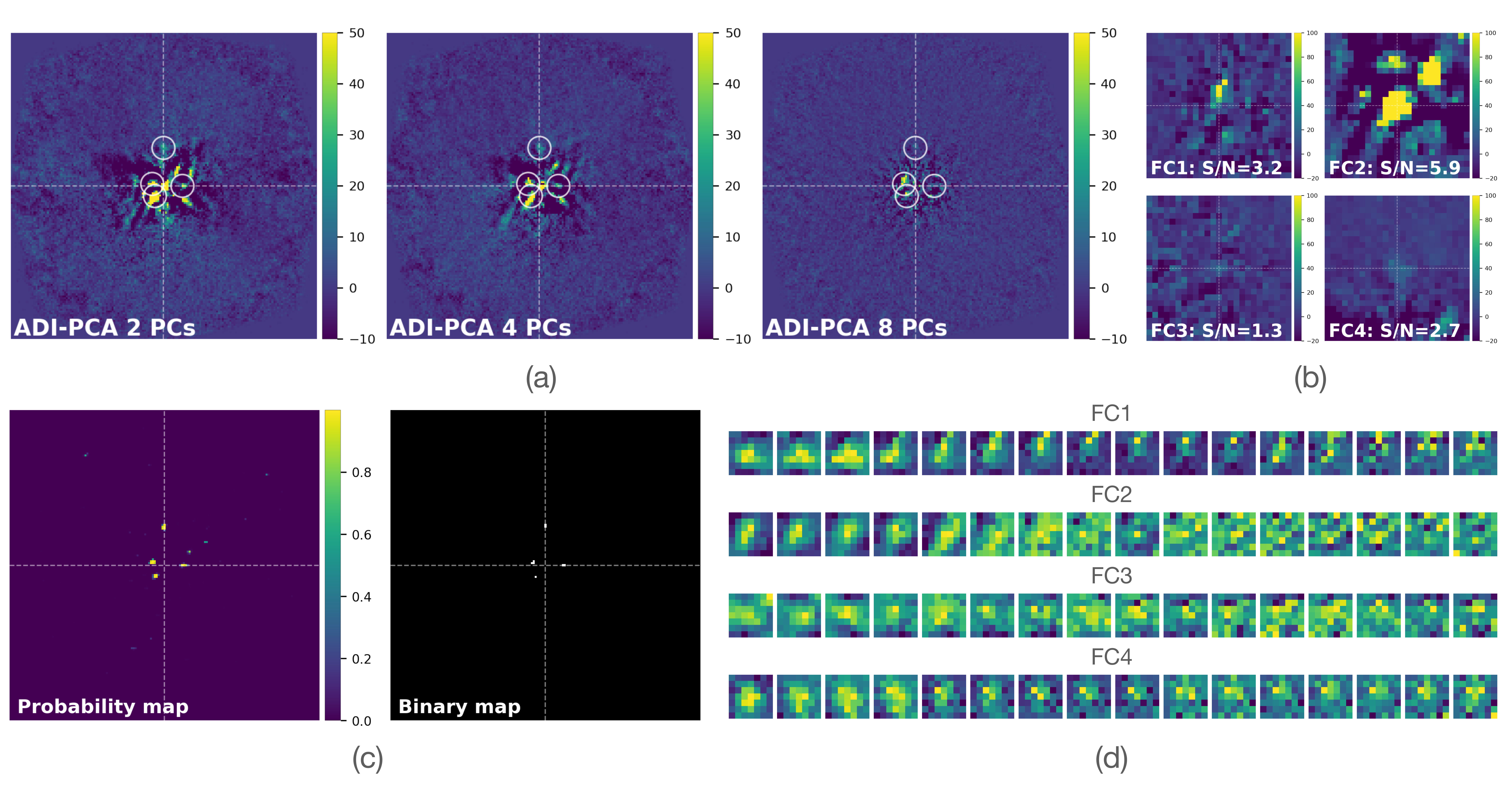}
	\caption{Injection of four synthetic companions (parameters detailed in Table \ref{tab:fctest}) in the V471 Tau VLT/SPHERE ADI sequence. The locations of the injections are shown with white circles on the ADI-PCA residual images. Panel (a) shows three ADI-PCA final frames with with two, four and eight PCs subtracted. Panel (b) shows cropped frames centered on the injected companions after optimizing the number of PCs (as shown in Table \ref{tab:fctest}) to maximize the S/N of each companion. SODINN's probability and binary maps clearly reveal the four planets (without false positives at 99\% probability) as seen in panel (c). Panel (d) shows the MLAR patches, used at the prediction stage, centered on each one of the injected companions.}
	\label{fig:flux_fctest}
\end{figure*}

\section{Performance assessment} \label{sec:perf_eval}
Testing on known companions is a first sanity check for any exoplanet detection algorithm. Next step is to proceed with testing the performance (detection capacity) of our trained models by injecting fake companions. In this section, we focus on SODINN. Using the V471 Tau VLT/SPHERE dataset, a challenging ADI sequence with few frames and mild rotation, we inject four companions (using the input off-axis PSF), at angular separations ranging from one to five $\lambda/D$, as indicated in Table \ref{tab:fctest} and illustrated in panel (a) of Fig \ref{fig:flux_fctest} (which shows three realizations of an ADI-PCA residual frame with two, four and eight PCs subtracted). The first, third and fourth companions are pretty much at the level of the speckle noise at their corresponding separations. The shapes of their PSFs are hard to distinguish from surrounding noise and the S/N values are small. The quoted S/N in panel (b) of Fig.~\ref{fig:flux_fctest} is the best mean S/N, in a 1$\times$FWHM aperture centered at the injection positions, obtained after optimizing the number of PCs (shown in Table \ref{tab:fctest}).
Only the second companion has a S/N over five, which is due to the fact that it was purposely injected on top of a bright speckle. The visual inspection would not be definitive for such a companion. As shown in panel (c) of Fig.~\ref{fig:flux_fctest}, SODINN outperforms the full-frame ADI-PCA approach by recovering the four companions at a high (99\%) probability without any false positive. 

\begin{table}
	\centering
	\caption{Parameters for the fake companions (FC) of Fig. \ref{fig:flux_fctest}.}
	\label{tab:fctest}
	\begin{tabular}{cccccc} 
		\hline
		\hline
		FC & Separation & PA & Flux(ADUs) & Contrast & PCs \\
		\hline
		1 & 1.5 $\lambda/D$ & 170$\degree$ & 9000 & \num{2.5e-4} & 8\\
		2 & 1.75 $\lambda/D$ & 230$\degree$ & 7000 & \num{1.9e-4} & 2\\
		3 & 2.5 $\lambda/D$ & 0$\degree$ & 1500 & \num{4.2e-5} & 9\\
		4 & 5 $\lambda/D$ & 90$\degree$ & 400 & \num{1.1e-5} & 4\\
		\hline
	\end{tabular}
\end{table}

\begin{figure}
	\includegraphics[clip,trim=60 100 120 90, width=9cm]{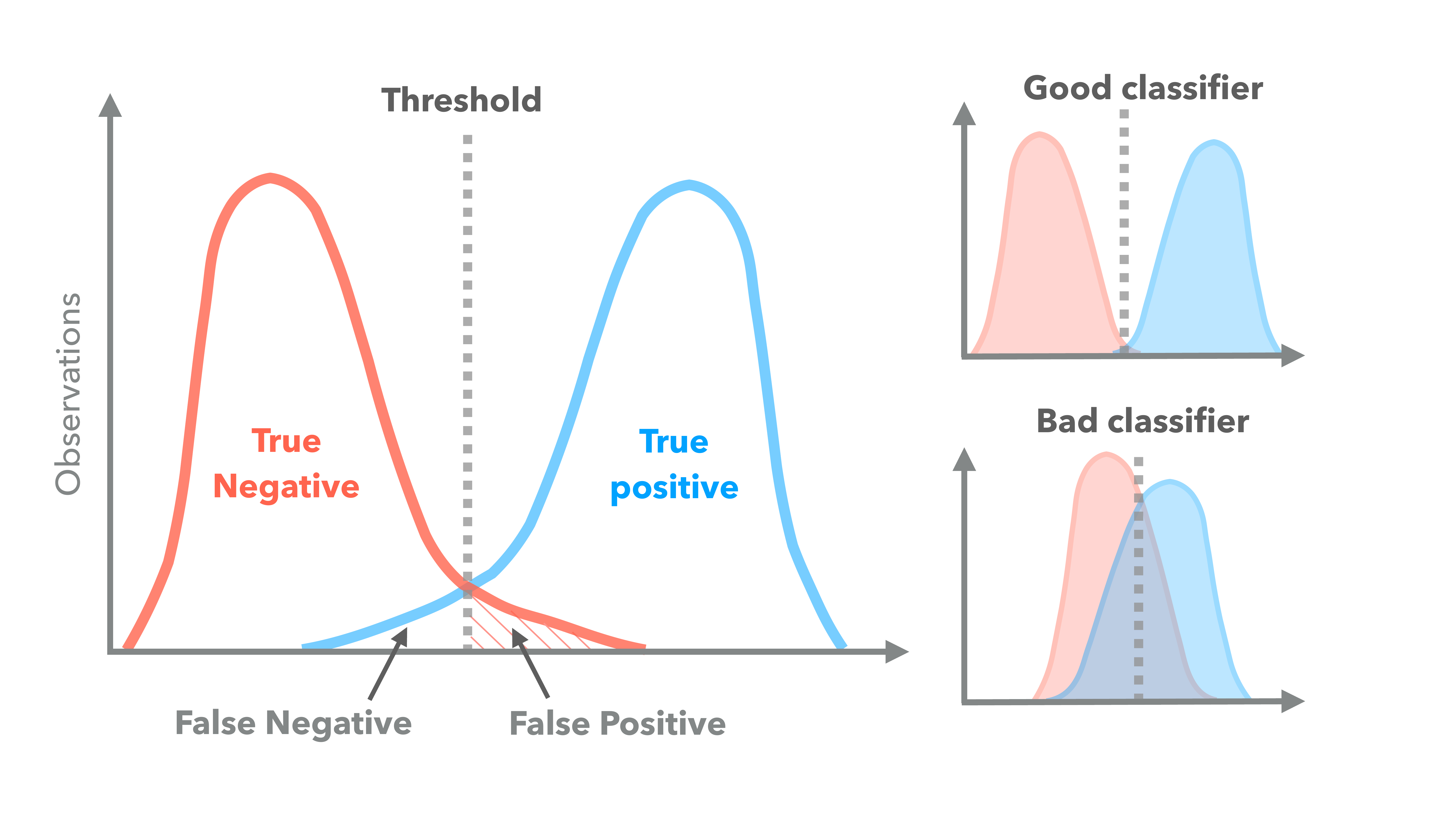}  
	\caption{Behavior of a binary classifier in a signal detection theory context. By varying the detection threshold we can study the classifier's performance.}
	\label{fig:binclass}
\end{figure}

Tests with known and injected companions are the first attempts to measure the performance of our supervised detection method. Unfortunately, it is not possible to judge the performance of a detection algorithm based on a few realizations of such tests. 
Following \citet{gomez16}, we use a robust signal detection theory tool for assessing the performance of our exoplanet detection algorithms: the receiver operating characteristic (ROC) curve. This curve is a graphical plot used for assessing the performance of classifiers (see Appendix \ref{app:roc} for a more detailed discussion). In general, ROC curves allow us to study the performance of a binary classifier system in a true positive rate (TPR $=p(\hat{y}=c^+ \mid y=c^+$)) - false positive rate (FPR $=1-p(\hat{y}=c^- \mid y=c^-$)) space, as a detection threshold $\tau$ varies. In other words, they can assess the TPR (also called sensitivity) and the FPR at the same time. 
In Fig.~\ref{fig:binclass}, we illustrate the task of a binary classifier in a signal detection context and the effect of choosing a detection threshold. By varying this threshold, we can adjust the FPR that we are willing to accept for a specific sensitivity. A ROC curve shows how good is our classification algorithm for separating the two classes, an ability inherent to the classifier. HCI as a signal detection problem seeks to simultaneously maximize the sensitivity to companions and minimize the number of false detections (FPR). 

\begin{figure}
	\centering
	\includegraphics[clip,trim=30 130 20 20,width=7.8cm]{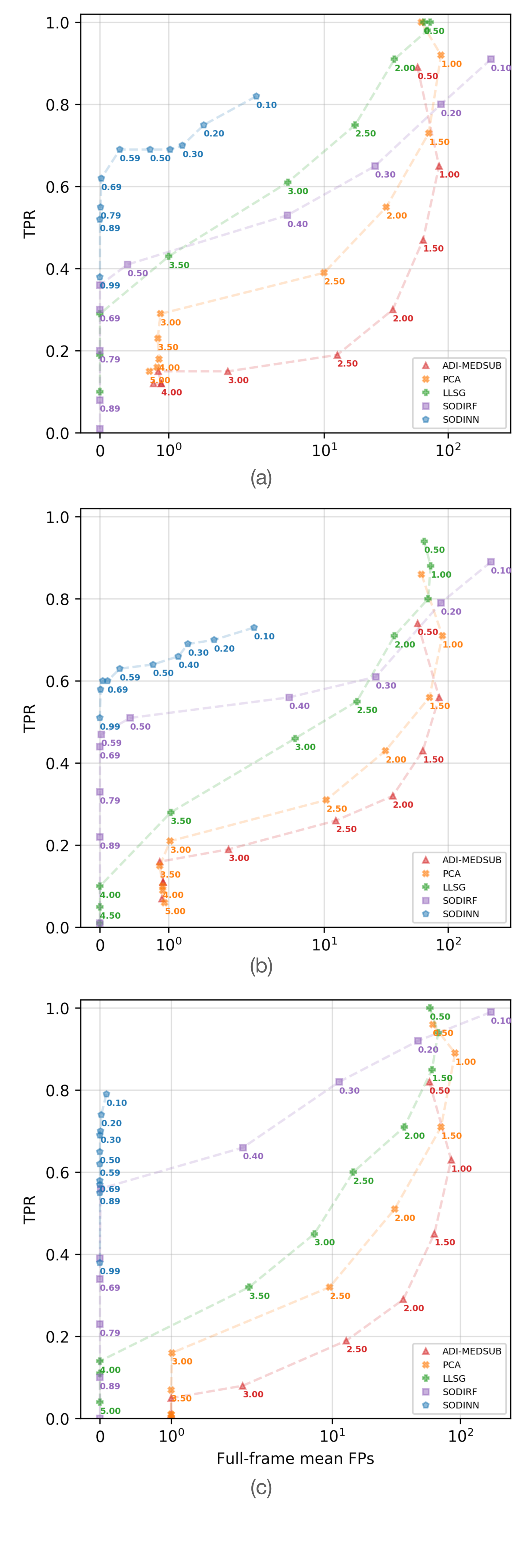}
	\caption{ROC curves for the VLT/SPHERE V471 Tau dataset, comparing ADI median subtraction, ADI-PCA, LLSG, SODIRF and SODINN. The panels show ROC curves built for different separations: (a) 1-2 $\lambda/D$, (b) 2-3 $\lambda/D$ and (c) 4-5 $\lambda/D$. The contrasts are shown in Table \ref{tab:roc_table}.}
	\label{fig:roc_curves}
\end{figure}

\begin{figure}
	\centering
	\includegraphics[clip,trim=30 130 20 20,width=7.8cm]{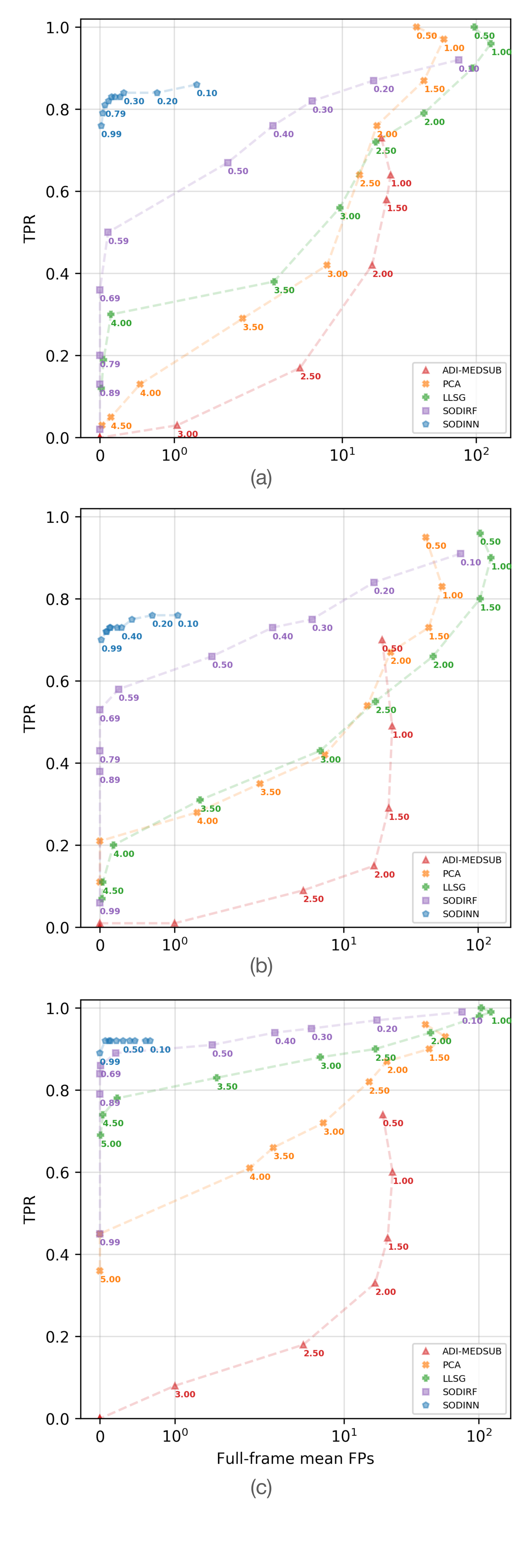}
	\caption{Same as Fig. \ref{fig:roc_curves} for the VLT/NACO $\beta$~Pic dataset. The contrasts are shown in Table \ref{tab:roc_table}. The labels denote the detection thresholds: S/N for ADI median subtraction, ADI-PCA and LLSG, and probabilities for SODIRF and SODINN.}
	\label{fig:roc_curves_bpic}
\end{figure}

In this study, we choose to build our ROC curves in a TPR (percentage of detected fake companions) vs mean per-frame false positives, instead of a TPR vs FPR space. The total number of false positives is counted on the whole detection map, and is averaged for each $\tau$. This reflects better the goal of a planet hunter and facilitates interpretation of the performance simulations. The ROC curves are built separately for different annuli with a tuned uniform flux distribution for the injection of fake companions. Having ROC curves for different separations from the star better illustrates the algorithm performance at different noise regimes. When interpreting the results, it is important to compare the ROC curves for different algorithms to each other, for a given annulus, considering that the TPR depends on the brightness of the injected companions, while the mean per-frame false positives does not (see panels (b) and (c) of Fig.~\ref{fig:app_rocs}). It is also important to examine the shape of the curves. For instance, it is preferable to have a steeper curve, which means that such algorithm does better in minimizing the number of FP while it increases its sensitivity.

We compare SODINN and SODIRF to classical ADI median subtraction, full-frame ADI-PCA and LLSG on both the VLT/NACO $\beta$~Pic dataset and the VLT/SPHERE V471 Tau dataset. As mentioned earlier, differential imaging approaches (unsupervised learning), i.e. ADI median subtraction, ADI-PCA and LLSG, do not generate a prediction (probability) but rather a residual image to look at. We obtain detection maps for these approaches by building S/N maps and thresholding them at several values of $\tau$. For each injection of a fake companion, a new data cube is built and processed with each of the five algorithms. In the case of the VLT/SPHERE V471 Tau dataset, the labeled datasets used for training SODIRF/SODINN are produced using both H2 and H3 SPHERE/IRDIS image sequences, while the prediction step is performed on the H3 band sequence only. The discriminative models are trained once for the ROC curve analysis. The number of PCs for ADI-PCA and the rank parameter for LLSG are set to two PCs (0.7 CEVR) for the V471 Tau sequence and to nine PCs (0.9 CEVR) for the $\beta$~Pic one. They are optimized in order to have the best possible ROC curves for ADI-PCA at the considered separations. No other hyperparameters were tuned. S/N maps were built for the resulting residual frames and thresholded at different values of $\tau$: 0.5, 1, 1.5, 2, 2.5, 3, 3.5, 4, 4.5, 5. For SODINN and SODIRF, we thresholded the probability map at several levels: 0.1, 0.2, 0.3, 0.4, 0.5, 0.59, 0.69, 0.79, 0.89, 0.99. Fig.~\ref{fig:app_detmapsall} illustrates one single realization of a companion injection, the generation of detection maps and the thresholding operation for three values of $\tau$. When training SODIRF and SODINN, MLAR samples of 16 slices (in the interval 0.5-0.95 CEVR) are used for the V471 Tau dataset, and 20 slices (in the interval 0.46-0.98 CEVR) for the $\beta$~Pic sequence.

\begin{table*}
	\centering
	\caption{Parameters used for the ROC curves of Figs. \ref{fig:roc_curves} and \ref{fig:roc_curves_bpic}.}
	\label{tab:roc_table}
	\begin{tabular}{c|c|c|c|c|c} 
		\hline
		Panel & Separation & V471 Tau, flux(ADUs) & V471 Tau, contrast & $\beta$~Pic, flux (ADUs) & $\beta$~Pic, contrast\\
		\hline
		(a) & 1-2 $\lambda/D$ & U(3000,7000) & \num{8.5e-5} to \num{1.9e-4} & U(400,900) & \num{5.2e-4} to \num{1.2e-3} \\
		(b) & 2-3 $\lambda/D$ & U(1000,5000) & \num{2.9e-5} to \num{1.4e-4} & U(50,450) & \num{6.5e-5} to \num{5.9e-4} \\
		(c) & 4-5 $\lambda/D$ & U(250,650) & \num{7.1e-6} to \num{1.8e-5} & U(10,210) & \num{1.3e-5} to \num{2.7e-4} \\
		\hline
	\end{tabular}
\end{table*}

The ROC curves, built for three different annuli, are shown in Figs. \ref{fig:roc_curves} and \ref{fig:roc_curves_bpic}. Brightnesses, contrasts and distances, for all the injected companions (100 for each annulus), are shown in Table \ref{tab:roc_table}. Reading the ROC curves presented here is straightforward: panel (a) of Fig. \ref{fig:roc_curves} (annulus from one to two $\lambda/D$) shows that a blob, i.e. at least two active pixels inside a 3$\times$3 pixels box centered at the position of the fake companion injection, sticks out above the detection threshold in 16\%, 28\%, $\sim$42\%, $\sim$44\% and $\sim$68\% of the cases for ADI median subtraction, ADI-PCA, LLSG, SODIRF and SODINN respectively, and for an average of $\sim$0.8 false positives in the full-frame detection map. The ROC curves for different separations and two very different datasets (from different HCI instruments) consistently show SODINN's improved performance with respect to other approaches. SODIRF's sensitivity improves with the separation and starts to match the performance of SODINN. In Appendix \ref{app:roc}, we provide more details about the construction of the ROC curves for the assessment of exoplanet detection algorithms. For instance, we show that hyperparameter tuning is important and the curves for ADI-PCA and LLSG could be slightly improved by searching the optimal number of PCs at each separation.

\section{Conclusions} \label{sec:conclusions}

This study illustrates the potential of machine learning in HCI for the task of exoplanet detection. We present a novel paradigm for detecting point-like companions in ADI sequences by reformulating HCI post-processing as a supervised learning problem, building on well-established machine learning techniques. Instead of relying on unsupervised learning techniques, as most of the state-of-the-art ADI post-processing algorithms do, we generate labeled datasets (MLAR samples) and train discriminative models that classify each pixel of the image, assigning a probability of containing planetary signal. We present two approaches that differ in the type of discriminative model used: SODIRF and SODINN. The former employs a random forest classifier while the latter features a more advanced deep neural network model, which exploits better the structure of the labeled MLAR samples. 

In order to assess the detection capabilities of our approaches, we perform a ROC analysis comparing both SODINN and SODIRF to ADI median subtraction, ADI-PCA and LLSG techniques. The performances of both algorithms are beyond what ADI-PCA and ADI median subtraction can offer. SODIRF can be considered as a computationally cheap alternative to the deep neural network approach of SODINN, whose performance lies in a separate zone of the ROC space. From one to two $\lambda/D$, SODINN improves the TPR by a factor of $\sim$2 and $\sim$10, for two different datasets, with respect to ADI-PCA and LLSG when working at the same false positive level. Moreover, the improvement in discriminating planet signal from speckles holds in the case of a challenging ADI sequence, with mild rotation and few frames, from a last-generation HCI instrument -- VLT/SPHERE (see Appendix \ref{app:roc} for a deeper discussion of the ROC curves performance assessment). The fact that these models are versatile and can be fine-tuned to each specific ADI sequence opens great possibilities of re-processing existing databases, from first- and second-generation HCI instruments, to maximize their scientific return. 

Although in this study we only addressed single ADI datasets, our framework's true potential is in the context of surveys, where the data from different observations could be used to generate a larger and more diverse labeled datasets. This would enable more efficient and general deep neural network models for SODINN. The exploitation of SODINN for surveys will be the focus of a future study. Other interesting venues of future research are the inclusion of the companion brightness into the model, the extension to other HCI observing techniques (beyond ADI), and the use of generative neural networks for complementing the data augmentation process. 

The simultaneous increase in sensitivity, which translates in deeper detection limits (the ability to detect companions at higher contrasts), and reduction of the per-image false positives clearly indicate that our supervised approach SODINN is a very powerful HCI exoplanet detection technique. Considering that ADI remains the most common HCI observing strategy and the large reservoirs of archival data, SODINN could potentially improve the demographics of directly imaged exoplanets at all separations, including those in the inner vicinity (1-2 $\lambda/D$) of their parent stars where ADI signal self-subtraction and speckle noise are the strongest.

\begin{acknowledgements}
The authors would like to thank the \texttt{python} open-source scientific community and the developers of \texttt{Keras} deep learning library. The authors acknowledge fruitful discussions and ideas from the participants in the Exoplanet Imaging and Characterization workshop organized by the W.M. Keck Institute for Space Studies. The research leading to these results has received funding from the European Research Council Under the European Union’s Seventh Framework Program (ERC Grant Agreement n. 337569) and from the French Community of Belgium through an ARC grant for Concerted Research Action.   
\end{acknowledgements}

\bibliographystyle{aa}   
\bibliography{biblio}    

\begin{appendix} 

\section{Construction of ROC curves} \label{app:roc} 

ROC curves are commonly used statistical tools for assessing the performance of binary classifiers. The planet detection task, where we are interested in evaluating the algorithm's sensitivity or ability to detect planets of varying contrast (brightness with respect to the star), can be seen as a binary classification. Therefore, ROC curves can be used for algorithms performance assessment in HCI \citep{barret06, lawson12}. A ROC curve shows a classifier TPR-FPR trade-off as a function of a detection threshold $\tau$. 

It is important to understand that the relative ROC performance of two different algorithms changes due to several factors: the dataset used (which has a set of characteristics such as the total rotation range, integration time, total number of frames, weather condition, wavefront control system performance and coronagraphic solution), hyper-parameter tuning of each algorithm (as shown in Fig.~\ref{fig:app_rocs}), noise regime or separation from the star, and contrast of the injected companions (as shown in Fig.~\ref{fig:app_rocs}). There is no shortcut to avoiding the dependence on these factors, unless the metric makes strong assumptions about the data and noise distributions (which are rarely confirmed in practice). A data-driven approach to the calculation of ROC curves, using standardized datasets, is the most fair and reliable method for assessing the performance of HCI algorithms. The ROC curves shown in this work, for the case of a single ADI dataset, are generated in the following way:
	
\begin{figure}
	\centering
	\includegraphics[clip,trim=20 60 20 0,width=6.5cm]{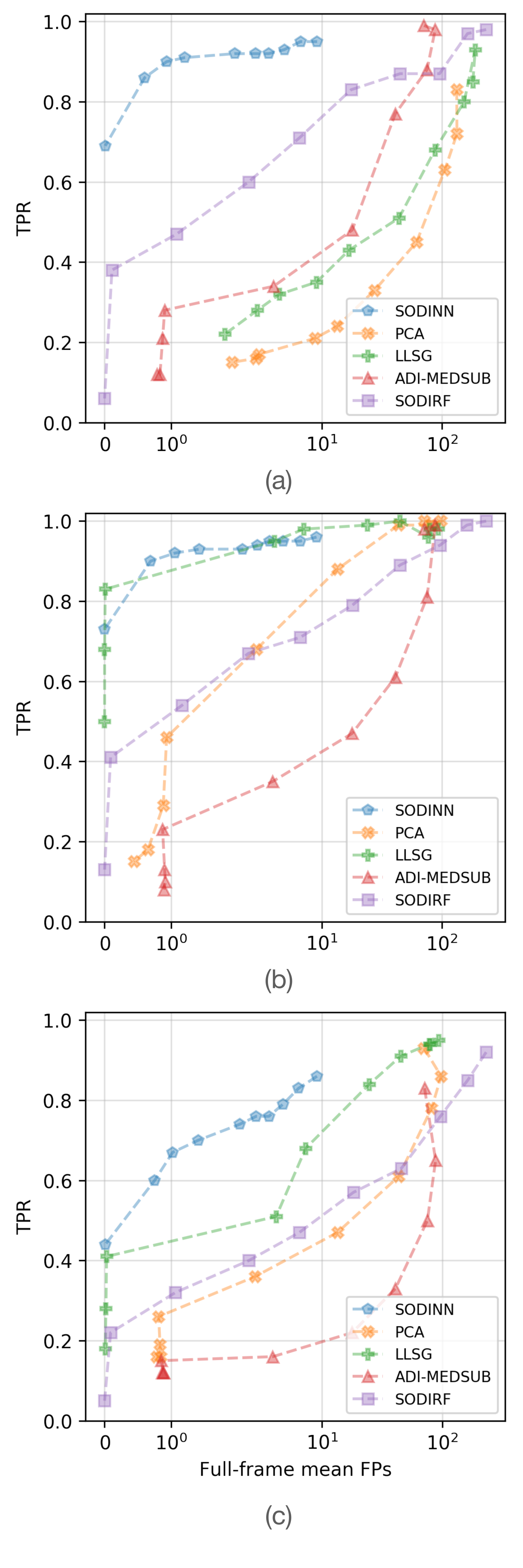} 
	\caption{This exemplifies the pitfalls of comparative studies using ROC curves, and how easy it is to obtain wrong relative performances and present unfair conclusions. These ROC curves are built for the same dataset and separation from the star. Panels (a) and (b) show ROC curves when changing the algorithms hyper-parameters: the number of PCs for ADI-PCA and the rank of LLSG. In (a) a more aggressive value is used with respect to (b). 
	The performance of ADI-PCA and LLSG is worst when too aggressive hyper-parameters are used. Notice how their curves move upward in panel (b) with respect to ADI median subtraction, SODIRF and SODINN curves. Panel (c) is generated injecting fainter companions with respect to panel (b). A higher planet to star contrast interval is a more sensible choice for highlighting the relative sensitivity of the studied algorithms. } 
	\label{fig:app_rocs}
\end{figure}
	
\begin{figure*}
	\centering
	\includegraphics[clip,trim=0 0 0 0, width=17cm]{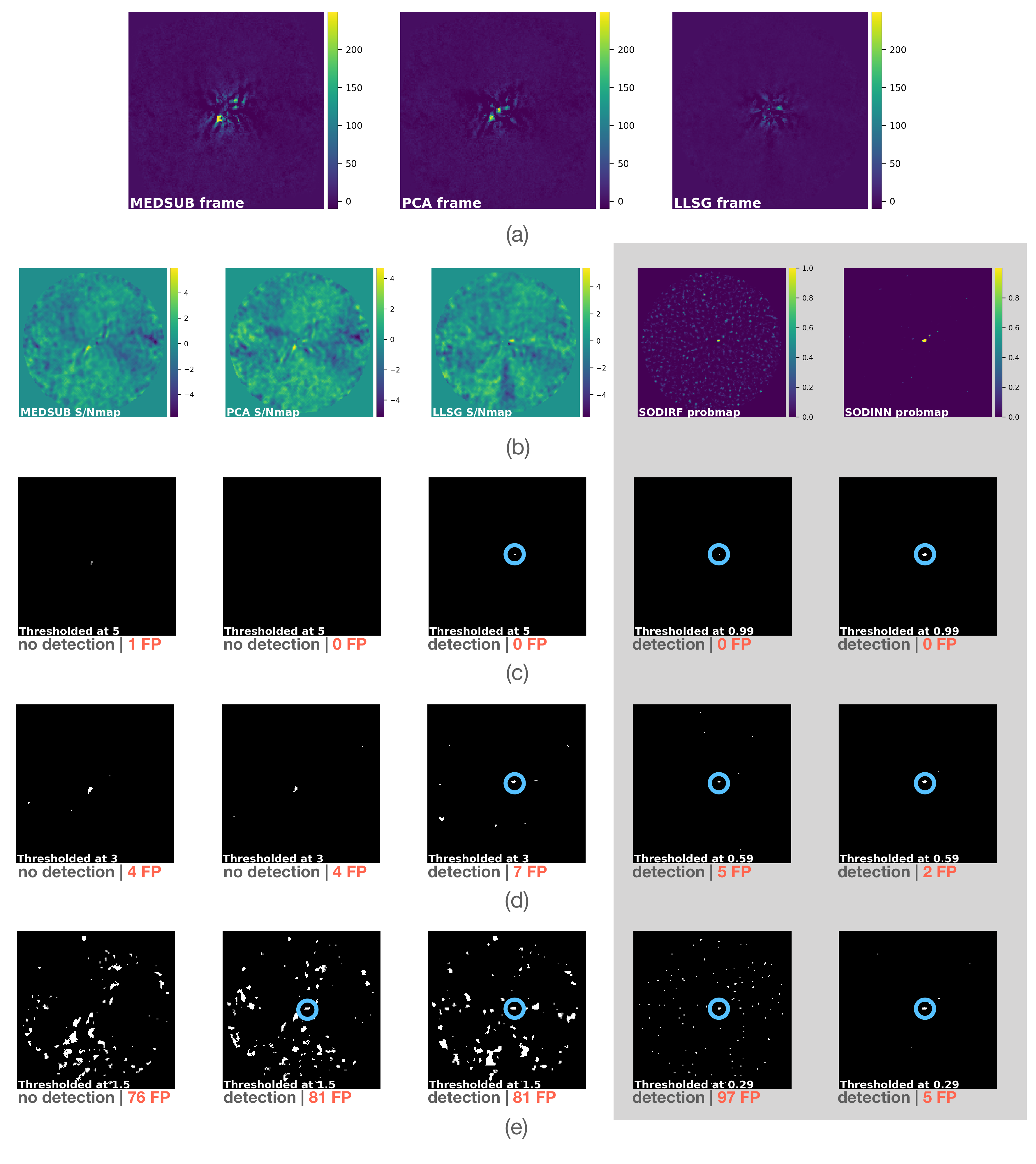}
	\caption{Case of a single injection for building a ROC curve comparative analysis. Panel (a) groups the final residual frames for the model PSF subtraction approaches (ADI median subtraction, ADI-PCA and LLSG). Detection maps are shown in panel (b): S/N maps from the residual flux frames of panel (a) and probability maps of SODIRF and SODINN. Panels (c), (d) and (e) show the binary maps obtained from the thresholded S/N and probability maps of panel (b). The detected fake companion is shown with a blue circle on the binary maps. The detection state and the number of FPs are also shown next to each binary map. Notice that the number of FPs grows when $\tau$ is decreased and also that SODINN controls the number of FPs. A large number of these injections (with varying flux and position) need to be performed in order to build the ROC curves.}
	\label{fig:app_detmapsall}
\end{figure*}

\begin{enumerate}
	\item An on-sky dataset is chosen. Any high-S/N or known companion is removed, e.g.  using the negative fake companion technique \citep{lagrange10,marois06adi,gomez17}.   
	\item A separation from the star (1$\times$FWHM annulus) and a planet to star contrast interval (the brightness of the injected companions) are selected. A list of $\tau$ thresholds is also defined.
	\item A large enough number of data cubes are built with a single injected companion at the selected separation and within the chosen contrast interval.
	\item The data cubes are processed with each algorithm involved in the performance assessment/comparison. Panel (a) of Fig.~\ref{fig:app_detmapsall} shows the resulting residual flux frames for the model PSF subtraction approaches. Panel (b) shows the resulting probability maps of SODIRF and SODINN. S/N maps are produced from the residual flux frames (see panel (b) of Fig.~\ref{fig:app_detmapsall}).
	\item Binary maps are obtained by thresholding the S/N and probability maps for different values of $\tau$ (see panels (c), (d) and (e) of Fig.~\ref{fig:app_detmapsall}). For each detection map and for each $\tau$, a true positive is counted if a blob is recovered at the injection location. False positives are other significant blobs at any other location in the detection map.
	\item For each $\tau$, the true positives and the number of false positives are averaged.
\end{enumerate}
	
When choosing a dataset, we must subtract known and high-S/N existing companions, based on visual vetting performed on a model PSF-subtracted residual image. As shown in this study, the PSF subtraction methods combined with visual vetting and S/N metrics are far from obtaining 100\% probability of finding companions and therefore obtaining an empty dataset. Nevertheless, the only choice is to assume the sequence is empty, or free of astrophysical exoplanetary signal, and flag any potential companion as a false positive in the following steps of the ROC curve generation procedure. In the last step, averaging the number of false positives (instead of assuming a static noise realization per $\tau$) addresses small fluctuations in this value, caused by the interaction of an injected companion with the false positives at the same separation (which biases the S/N).
	
\end{appendix}

\end{document}